%% file: main.tex
\documentclass[a4paper]{article}

\usepackage{amssymb}

\newcommand{\Z}{\mathbb{Z}}

\newcommand{\C}{\mathbb{C}}

\newcommand{\Ob}{\mathcal{O}}
\newcommand{\J}{\mathcal{J}}
\newcommand{\X}{\mathcal{X}}
\newcommand{\A}{\mathcal{A}}
\newcommand{\D}{\mathcal{D}}
\newcommand{\F}{\mathcal{F}}

\def\ctimes{\stackrel{\otimes}{,}}

\newcommand{\tpoisson}[2]{\left\{#1 \ctimes #2\right\}}

\def\half{{1 \over 2}}
 
\newcommand{\transpose}[1]{{\vphantom{#1}}^{\mathit t}{#1}}

\def\operator@font{\mathgroup\symoperators}
\def\ch{\mathop{\operator@font ch}}
\def\tr{\mathop{\operator@font tr}}
\def\image{\mathop{\operator@font Im}}

\gdef\IncludeDetails{0}

\begin{document}

\rightline{LPTHE-02-17}
\vskip 2cm

\centerline{\LARGE Affine Jacobians of Spectral Curves}
\centerline{\LARGE      and Integrable Models         }
\vskip 1cm

\centerline{F.A.~Smirnov\footnote[0]{Membre du CNRS} and V.~Zeitlin}
\vskip 1cm

\centerline{Laboratoire de Physique Th\'eorique et Hautes Energies
            \footnote[1]{\it Laboratoire associ\'e au CNRS.}}
\centerline{Universit\'e Pierre et Marie Curie, Tour 16, 1$^{er}$ \'etage,
            4 place Jussieu}
\centerline{75252 Paris Cedex 05, France}
\vskip 1.5cm

\begin{abstract}
    We explicitly construct the algebraic model of affine Jacobian of a
    generic algebraic curve of high genus and use it to compute the Euler
    characteristic of the Jacobian and investigate its structure.
\end{abstract}

\vskip 1cm

\input intro.tex
\input curve.tex

\input brackets.tex

\input separation.tex

\input reconstr.tex

\input euler.tex

\input end.tex

\end{document}

%% file: intro.tex
\section{Introduction}

\subsection{Summary}

The goal of this article is to expand the constructions presented in
\cite{NScohom} for the case of hyperelliptic curves (case $N=2$) to the
general spectral curves of any order.

As in \cite{NScohom} our main tool is the affine model of the Jacobi variety
of the curve or, in the other words, the construction of the separated
variables for the underlying integrable system. Although the existence of such
construction has already been proven in \cite{Beauville} it only becomes
really useful in the study of the integrable systems if it can be specified
explicitly. It has been known for a long time for the hyperelliptic case (see
\cite{Mumford} for example) and was originally done by Sklyanin in
\cite{Sklyanin} for $SL(3)$. His ideas based on the functional Bethe Ansatz
were later generalized to $SL(N)$ in \cite{Scott} while a more elegant
algebraic-geometrical approach (which is the closest to our one) was used in
\cite{Dubrovin}.

However neither of these models could be used to generalize the results of
\cite{NScohom} because we need to choose a particular orbit carefully (see
the section \ref{sec:poisson}) and only then the construction of the separated
variable gives us the additional insights into the structure of the cohomology
groups.

With this problem behind us, the other results obtained for hyperelliptic
curves carry over to the general case with only technical problems. But in
other respects the general case is quite different and almost surely much more
complicated. Because of this, we are still unable to even conjecture the
explicit form of cohomologies for it. We do show that the highest nontrivial
cohomology group doesn't suffice any more to generate the elements of the ring
of observables $\A$ for $N>2$ in a marked difference with the hyperelliptic
case.

\subsection{The Plan}

In the first section after this one we introduce the spectral curve $\Sigma$
which is the starting point of our study. In the context of integrable
systems, this curve is the phase space of some integrable model and the
the ring $\A$ of meromorphic functions on the Jacobian of this curve with
singularities only on the theta divisor corresponds to the ring of
observables.

In the section \ref{sec:poisson} we construct the matrix space which will be
later shown to provide an algebraic model of $\A$ and supply it with the
Poisson structure. This task is quite difficult in the general case as it is
less obvious to identify the `constant' ring elements (in the hyperelliptic
case we only had to deal with the center of mass motion one) which must not be
taken into account if we are to get the isomorphism with the observables ring.

Next, in the section \ref{sec:separation}, we construct the separated
variables which provide us with the mapping from the matrices lying in the
space discussed above to the point on the Jacobian of $\Sigma$.

Section \ref{sec:reconstr} deals with a much more difficult problem of
constructing the inverse mapping. Although it is not as explicit as in
hyperelliptic case, we still were able to find a constructive way to express
it. The existence of this mapping allows us to conclude that our space of
matrices is indeed isomorphic to the affine Jacobian.

Finally, we combine the preceding results together in the section
\ref{sec:euler} to calculate the Euler characteristic of the Jacobian and
discuss the cohomologies structure which would be compatible with it. As
indicated above, we are unfortunately forced to conclude that the simplest
situation encountered in studying the hyperelliptic case doesn't hold in
general.

%% file: curve.tex
\section{The Spectral Curve}

\subsection{General properties}

We consider an algebraic curve $\Sigma$ of genus $g$ and its theta divisor
$\X$ which is a $(g-1)$-dimensional subvariety of the Jacobi variety $\J$
associated to $\Sigma$. We call the affine Jacobian of $\Sigma$ the
non-compact variety $\J - \X$.

Our construction of the algebraic model of the affine Jacobian is only
applicable to the spectral curves, that is curves defined by the equation
\begin{equation}
    r(w, z) \equiv w^N + t_1(z)w^{N-1} + \ldots + t_N(z) = 0
    \label{eq:curve}
\end{equation}
where $t_k(z)$ are polynomials in $z$ of degree at most $kn - 1$ and the
degree of $t_N(z)$ is exactly $Nn-1$.

The genus of the curve (\ref{eq:curve}) is easily computed to be
\[
    g = \half (N-1)(Nn-2)
\]
and we can write down the abelian differentials of the first kind on it
explicitly as
\[
    \sigma_{kl}(w, z) \equiv \frac{w^l z^k}{\partial_w r(w, z)} d\,z
        \qquad \quad \hbox{with} \quad
        \left\{
            \begin{array}{l}
                0 \leq l < N-1 \\
                0 \leq k < (N - l - 1)n - 1
            \end{array}
        \right.
\]
(we shall also denote them below as $\sigma_I$ with a composite index $I=1,
\ldots, g$ instead of two indices $k$ and $l$ for convenience).

Also note that the second kind differentials on $\Sigma$ have the same
form except that $l$ and $k$ are unlimited for them.

Let us also choose a canonical homology basis of $\Sigma$ $(\alpha_1, \ldots,
\alpha_g, \beta_1, \ldots, \beta_g)$ and define the corresponding normalized
differentials $\omega_I$
\[
    \omega_I = \sum_{J=1}^g (A^{-1})_{IJ} \sigma_J
\]
where $A$ is the matrix of $\alpha$-periods:
\[
    A_{IJ} \equiv \oint_{\alpha_J} \sigma_I
\]

These differentials are normalized because we have
\[
    \left\{
        \begin{array}{lcl}
            \oint_{\alpha_I} \omega_J & = & \delta_{IJ} \\
            \oint_{\beta_I} \omega_J  & = & B_{IJ}
        \end{array}
    \right.
\]
where $B$ is the period matrix (it is symmetric, with positive definite
imaginary part).

The Jacobi variety $\J$ of $\Sigma$ is then simply defined as a
$g$-dimensional complex torus
\[
    \J = \frac{\C^g}{\Z^g + B \Z^g}
\]

The Riemann theta function associated with the Jacobian $\J$ is defined by
\[
    \theta(\xi) = \sum_{n \in \Z^g}
                    \exp 2 \pi i \transpose{n} (\half B n + \xi)
\]
It satisfies the quasi-periodicity condition
\[
    \theta(\xi + m + Bn) =
        \exp \left[-2 \pi i \transpose{n} (\half B n + \xi) \right] \theta(\xi)
\]
for any $m, n \in \Z^g$.

Thanks to the above we see that the $\theta$-divisor
\[
    \Theta = \left\{ \xi | \theta(\xi) = 0 \right\}
\]
is a (possibly singular) $(g-1)$-dimensional subvariety of $\J$.

\subsection{Holomorphic functions on the affine Jacobian}

From the point of view of the integrable systems theory, the observables should
be the holomorphic functions on the affine Jacobian or, in other words,
the meromorphic functions on $\J$ with singularities on the theta divisor
$\Theta$ only.

Such functions may be described using the products and quotients of the theta
functions themselves but for our purposes an alternative description is more
convenient.

First, let us introduce the Abel map. Note that the curve (\ref{eq:curve}) has
a single point at infinity which we denote by $\infty$ as in local coordinates
$z'=1/z$ and $w'=w/z^N$ the curve equation takes form $w'^N + z'(\ldots) = 0$.
Accordingly we may define the Abel map using $\infty$ as the base point:
\[
    \begin{array}{rcl}
        A: \Sigma^{\times g} & \rightarrow & \J \\
        \left(p_1, \ldots, p_g\right)
            & \mapsto
            & \sum_{k=1}^g \int_{\infty}^{p_k} \sigma + \Delta(\infty)
    \end{array}
\]
where $\Delta$ is the Riemann characteristic (a vector of constants depending
only on $\Sigma$ and the choice of the base point). The map $\A$ is
well-defined as deforming the integration path by a loop $\alpha_l$ only
changes the $k$-th component of the RHS by $\delta_{kl}$ while making an extra
$\beta_l$ loop adds $B_{kl}$ to it and thus deforming the path in any way
doesn't change the point on the Jacobian.

By a corollary of Riemann's theorem any $\xi \in \Theta$ may be written as
\[
    \xi = \sum_{j=1}^{g-1} \int_{\infty}^{q_j} \sigma + \Delta(\infty)
\]
If we write $\xi$ as an image of a point of $\Sigma^{\times g}$ by the Abel
map:
\[
    \xi = A(p_1, \ldots, p_g)
\]
we see that we have two equivalent divisors:
\[
    \left( p_1, \ldots, p_g \right) \simeq
        \left( q_1, \ldots, q_{g-1}, \infty \right)
\]
or, in other words, there exists a function with the simple poles in $p_i$ and
simple zeroes in $q_j$ and $\infty$ (we can always shift it by the value in
$\infty$ to ensure that it has a zero there).

If all $p_i$ are distinct, to find a function with the singularities on
$\Theta$ we have to find a function with simple zeroes in $(p_i)$. This is
possible if and only if
\[
    \Delta(p_1, \ldots, p_g) \equiv \det\left(w_i(p_j)\right) = 0
\]

The situation is more complicated in the case when two (or more) points
coincide but we can cancel the zeroes due to this by considering functions of
the following form:
\begin{equation}
    \frac{\det\left(\tilde{\sigma}(p_j)\right)}{\Delta(p_1, \ldots, p_g)}
    \label{eq:fholo}
\end{equation}
where $\tilde{\sigma}$ are any differentials of first or second kind.

Thus in any case the functions of the form (\ref{eq:fholo}) have poles on the
theta divisors only. Moreover, by reversing the argument above, we also see
that we can get all holomorphic functions on the affine Jacobian in the form
of products of functions (\ref{eq:fholo}).

%% file: brackets.tex
\def\too{t^{00}}
\def\tmp{t^{-+}}
\def\tpm{t^{+-}}

\section{The Affine Model}
\label{sec:poisson}

\subsection{The Matrix Poisson Brackets}

Consider a space of $N \times N$ matrices of polynomials of $z$:
\[
    M(z) = \left(m_{ij}(z)\right)_{1 \leq i,j \leq N}
\]
We start with the usual expression for the Poisson brackets for such
matrices:
\begin{equation}
    \left\{m(z_1) \ctimes m(z_2)\right\} =
        \left[ r_{12}(z_1,z_2), m(z_1) \otimes I + I \otimes m(z_2) \right]
\label{eq:brackets}
\end{equation}
where the $r$-matrix is defined by
\[
    r_{12}(z_1,z_2) = \frac{1}{z_1-z_2}\left(
                        \frac{1}{2}(z_1+z_2)t^{00} +
                        z_1 t^{-+} +
                        z_2 t^{+-}
                      \right)
\]
and
\begin{eqnarray*}
    t^{00} & = & \sum E_{ii} \otimes E_{ii} \\
    t^{+-} & = & \sum_{i<j} E_{ij} \otimes E_{ji} \\
    t^{-+} & = & \sum_{i>j} E_{ij} \otimes E_{ji}
\end{eqnarray*}
where the matrices $E_{ij}$ are $N \times N$ with the only non zero element
being $1$ at the intersection of $i$-th row and $j$-th column.

The brackets (\ref{eq:brackets}) are compatible with the following form of the
polynomials $m(z)$:
\begin{equation}
\left\{
\begin{array}{lcll}
    m_{ij}(z) & = & m^{(0)}_{ij} z^{n-1} + \ldots + m^{(n-1)}_{ij}
        & \quad \hbox{if } i \leq j \\
    m_{ij}(z) & = & m^{(0)}_{ij} z^n + \ldots + m^{(n-1)}_{ij} z
        & \quad \hbox{otherwise}
\end{array}
\right.
\label{eq:degm}
\end{equation}

With such form of the matrix elements we can easily see that
\begin{equation}
    \det m(z) = z^{Nn-1} \left(
                    m^{(0)}_{1N} \prod_{j=1}^{N-1} m^{(0)}_{j+1,j}
                \right) + \Ob(z^{Nn-2})
    \label{eq:detm}
\end{equation}
because only the minor containing the last element of the first row and
diagonal below the main one contributes the $(Nn-1)$-th power of $z$.

The characteristic polynomial of $m(z)$ has the following form:
\begin{equation}
    \det \left(m(z) + w\right) = w^N + t_1(z)w^{N-1} + \ldots + t_N(z)
    \label{eq:def_t}
\end{equation}
with coefficients $t_k(z)$ being polynomials of degree $k n - 1$ in $z$:
\[
    t_k(z) = z^{k n - 1}t_k^{(0)} + \ldots + t_k^{(k n - 1)}
\]
which means that the equation
\begin{equation}
    \det \left(m(z) + w\right) = 0
    \label{eq:curvem}
\end{equation}
defines a spectral curve in the sense of the previous section. We shall fix
the characteristic polynomial of $m(z)$ so that this equation describes the
same curve $\Sigma$ as the equation (\ref{eq:curve}).

All coefficients $t_k^{(i)}$ are in involution which follows from the identity
\begin{eqnarray}
    \lefteqn{ \left\{ \log \det (w + m(z_1)), m(z_2) \right\} = } \nonumber \\
        & = &
        \tr \otimes I \left( \left((w + m(z_1) \otimes I)\right)^{-1}
        \left[r_{12}(z_1, z_2), I \otimes m(z_2)\right] \right)
    \label{eq:brdet}
\end{eqnarray}
(which is easily proved by replacing $\log \det$ with $\tr \log$ and
developing the brackets of the logarithm).

Moreover, many of these coefficients lie in center of the Poisson algebra:
indeed, the coefficients $t_k^{(0)} \ldots t_k^{(n-1)}$ appear in front of
degrees of $z$ which are too big to appear in the RHS of (\ref{eq:brdet}).
In particular, the dominant coefficient of the determinant (\ref{eq:detm})
lies in the center.

\subsection{The `Physical' Matrix $l(z)$}

We cannot hope that $m(z)$ provide us with an algebraic model of the affine
Jacobian for several reasons.

First and the most obvious one is that we want to embed a curve of genus $g =
\half (N-1)(Nn-2)$ into the matrix space whose dimension is, as seen from
(\ref{eq:degm}), equal to $n N^2$. Choosing the subspace of matrices with the
fixed characteristic polynomial yields $n + 2n + \ldots + Nn = \half n N(N+1)$
constraints somewhat reducing the dimension of this subspace but we still have
\[
    n N^2 - \half n N(N+1) = \half n N(N-1) = g + (N - 1)
\]
and so we must get rid of $N-1$ extra `unphysical' degrees of freedom if we
are to establish an isomorphism.

There is another motivation for doing what follows which is less obvious but
even more important: in fact, it is impossible to reconstruct the full matrix
$m(z)$ in terms of the single-valued functions on the jacobian as it contains
some $\theta$-function like pseudo functions which cannot be expressed with
the help of holomorphic functions on (a subvariety of) $\J$. It also explains
why eliminating these extra degrees of freedom is far from being trivial:
their commutation relations with the other elements are quite involved (in
particular, they are not at all in the center of the Poisson algebra) and so
we cannot just remove them from consideration. Instead we have to go through
the construction below in order to obtain something that we can expect (and
actually will) to express in terms of functions on the jacobian.

To solve both of these problems, let us replace $m(z)$ with a conjugated
matrix
\[
    l(z) = s m(z) s^{-1}
\]
with $s$ built from $m(z)$ in the way described below:

We have, because of (\ref{eq:degm}),
\begin{equation}
    m(z) = z^n \mu^- + z^{n-1} \mu + \Ob(z^{n-2}) \qquad
                                                  \hbox{when } z \to \infty
    \label{eq:minfty}
\end{equation}
with $\mu^-$ being a strictly lower triangular matrix and $\mu$ -- an upper
triangular one (with diagonal terms). Then if we denote the first row of $\mu$
as $\mu_1$, we define the $N \times N$ matrix $s$ as
\[
    s = \left( \begin{array}{c}
                    \mu_1 (\mu^-)^{N-1} \\
                    \vdots \\
                    \mu_1 \mu^- \\
                    \mu_1
               \end{array}
        \right)
\]
(note that it is a lower triangular matrix -- simply because $\mu^-$ is).

Replacing $m(z)$ with $l(z)$ plays the role of `gauge fixing': indeed, $l(z)$
still defines the same curve (\ref{eq:curvem}) so the physical picture doesn't
change but all the extra degrees of freedom have disappeared from $l(z)$. To
see it, let us look at the degrees of the polynomials $l(z)_{ij}$.

First, we have:
\begin{eqnarray*}
    l(z) & = & s \mu^- s^{-1} z^n + \Ob(z^{n-1}) \\
         & = & U z^n + \Ob(z^{n-1})
    \label{eq:ldominant}
\end{eqnarray*}
where
\begin{equation}
    U = \left(
            \begin{array}{ccccc}
            0 & 0 & 0 & \ldots & 0 \\
            1 & 0 & 0 &        & 0 \\
            0 & 1 & 0 &        & 0 \\
            \vdots &   & \ddots & \ddots & \vdots \\
            0 & \ldots & \ldots & 1 & 0
            \end{array}
        \right)
    \label{eq:Umatrix}
\end{equation}
which shows that the degrees of all elements except the ones on the diagonal
below the main one are strictly less than $n$.

Next, let us determine the degrees of the elements in the first row. As
\begin{eqnarray*}
    (s \mu s^{-1})_{1j} & = & s_{11} (\mu s^{-1})_{1j} \\
                        & = & s_{11} (s s^{-1})_{Nj} \\
                        & = & \delta_{Nj} s_{11}
\end{eqnarray*}
we have
\[
    \left\{
        \begin{array}{lcll}
            l_{1j} & = & \Ob(z^{n-2}) & (j \leq N-1) \\
            l_{1N} & = & s_{11} z^{N-1} + \Ob(z^{n-2})
        \end{array}
    \right.
\]

So the degrees of the elements of $l(z)$ have the following form:
\begin{equation}
    \deg_z l(z) = \left(
                    \begin{array}{cccccc}
                        n-2 & n-2 & n-2 & \ldots & n-2 & n-1 \\
                        n   & n-1 & n-1 & \ldots & n-1 & n-1 \\
                        n-1 & n   & n-1 & \ldots & n-1 & n-1 \\
                        \vdots & & \ddots & \ddots & & \vdots \\
                        \vdots & & & \ddots & \ddots & \vdots \\
                        n-1 & \ldots & \ldots & \ldots & n & n-1
                    \end{array}
                  \right)
    \label{eq:ldegree}
\end{equation}

Note that $s_{11}$, which is the dominant coefficient of $l_{1N}$, is also
exactly the dominant coefficient of the (\ref{eq:detm}) and so it lies in the
center of the Poisson algebra or, in other words, is a constant.

Finally, the dominant coefficients of the other elements of $l(z)$ with
maximal degrees, that is the polynomials $l_{21}(z), \ldots, l_{N\,N-1}(z)$
are constants as well (as seen from (\ref{eq:ldominant}).

Calculating the number of coefficients in the matrix $l(z)$ using
(\ref{eq:ldegree}) (not counting the constant ones determined above) we see
that now we have exactly the `correct' number of parameters.

\subsection{Calculation of the Poisson Brackets of $l(z)$}

Our next goal is to calculate the Poisson brackets of $l(z)$. For this we
start by computing the brackets of $\mu$ and $m(z)$.

First, fix $z_2$ and let $z_1 \to \infty$. Then we have:
\ifnum\IncludeDetails=1
\begin{eqnarray*}
    \lefteqn{\tpoisson{m(z_1)}{m(z_2)} =} \\
    & = & {1 \over z_1 - z_2}
        \left[
            \half (z_1+z_2)t^{00} + z_1 t^{-+} + z_2 t^{+-},
            m(z_1) \otimes I + I \otimes m(z_2)
        \right] \\
    & = & \left[ \half t^{00} + t^{-+} + {z_2 \over z_1} \left(
                       t^{00} + t^{-+} + t^{+-}
                 \right),
                 m(z_1) \otimes  I
        \right]
\end{eqnarray*}
Or, with the help of (\ref{eq:minfty}) and setting
$t \equiv \too + \tmp + \tpm$:
\[
    \left\{\mu + z_1 \mu^- \ctimes m(z_2)\right\} =
        \left[\half t^{00} + t^{-+} + {z_2 \over z_1} t,
              (\mu + z_1 \mu^-) \otimes I
        \right]
\]
from where we directly see that
\fi
\begin{eqnarray}
    \tpoisson{\mu^-}{m(z_2)} & = &
        \left[ \half \too + \tmp, \mu^- \otimes I \right] \\ \nonumber
    \tpoisson{\mu}{m(z_2)} & = &
        \left[ \half \too + \tmp, \mu \otimes I \right] +
        z_2 \left[ t, \mu^- \otimes I \right]
    \label{eq:mubrackets}
\end{eqnarray}
\ifnum\IncludeDetails=0
with
\[
    t \equiv \too + \tmp + \tpm
\]
\fi

Next step is to develop $\left\{ l(z_1) \ctimes l(z_2) \right\}$ using that
$\left\{ s \ctimes s \right\} = 0$:
\begin{equation}
    \begin{array}{lcl}
    \tpoisson{l(z_1)}{l(z_2)} & = &
        (s \otimes s) \tpoisson{m(z_1)}{m(z_2)} (s^{-1} \otimes s^{-1}) + \\
        & & \left[ \kappa_{12}, l(z_1) \otimes I \right] -
            \left[ \kappa_{21}, I \otimes l(z_2) \right]
    \end{array}
    \label{eq:lbrackets1}
\end{equation}
where
\[
    \kappa_{12} = (I \otimes s)
                  \tpoisson{s}{m(z_2)}
                  (s^{-1} \otimes s^{-1})
\]

To compute $\kappa_{12}$ we first write
\[
    \tpoisson{s_j}{m(z_2)} = \tpoisson{\mu_1}{m(z_2)} \*
        \left(\left(\mu^-\right)^{N-j} \otimes I\right) +
        \left(\mu_1 \otimes I\right) \tpoisson{\left(\mu^-\right)^{N-j}}{m(z_2)}
\]
where $s_j = \mu_1 \left(\mu^-\right)^{N-j}$ is the $j$-th row of $s$. Using
(\ref{eq:mubrackets}) we have
\begin{eqnarray*}
    \lefteqn{ \tpoisson{s_j}{m(z_2)} = } \\
    & = & \half (\mu_1 \otimes E_{11})
          \left(\left(\mu^-\right)^{N-j} \otimes I\right) +
          z_2 \sum_k E_{jk} \left(\mu^-\right)^{N-j+1} \otimes E_{k1} - \\
    & - & (\mu_1 \otimes I)(\half \too + \tmp)
          \left(\left(\mu^-\right)^{N-j} \otimes I\right) + \\
    & + & (\mu_1 \otimes I)(\half \too + \tmp)
          \left(\left(\mu^-\right)^{N-j} \otimes I\right) - \\
    & - & (\mu_1 \otimes I) \left(\left(\mu^-\right)^{N-j} \otimes I\right)
          (\half \too + \tmp)
\end{eqnarray*}

Hence after simplification
\[
    \tpoisson{s}{m(z_2)} =
        (s \otimes I)(-\half \too - \tmp + \half I \otimes E_{11}) +
        z_2 \sum_{j,k} E_{jk} (\mu^-)^{N-j+1} \otimes E_{k1}
\]
And so
\begin{eqnarray*}
    \kappa_{12} & = & -(s \otimes s)
                   (\half \too + \tmp - \half I \otimes E_{11})
                   (s^{-1} \otimes s^{-1}) + \\
                &   & z_2 (I \otimes s)
                  \left(\sum_{jk} E_{jk}(\mu^-)^{N-j+1} \otimes E_{k1}\right)
                  (s^{-1} \otimes s^{-1})
\end{eqnarray*}

We can leave out the term with $I \otimes E_{11}$ as it commutes with
$l(z_1) \otimes I$. The second term can be simplified using the multiplicative
properties of $E_{rs}$:
\begin{eqnarray*}
    (1 \otimes s)\left(\sum_{jk} E_{jk}(\mu^-)^{N-j+1} \otimes E_{k1}\right)
        & = & \sum_{jkl} E_{jk}(\mu^-)^{N-j+1} \otimes s_{lk} E_{l1} \\
        & = & \sum_{jl} E_{jl} s (\mu^-)^{N-j+1} \otimes E_{l1} \\
        & = & \left(\sum_{jl} E_{jl} U^{N+1-j} \otimes E_{l1}\right)
              (s \otimes I)
\end{eqnarray*}
with the matrix $U$ defined in (\ref{eq:Umatrix}).

Finally, inject this expression into (\ref{eq:lbrackets1}) to obtain the
Poisson brackets of $l(z)$:
\begin{equation}
    \tpoisson{l(z_1)}{l(z_2)} =
        \left[\hat{r}_{12}(z_1,z_2),l(z_1) \otimes I\right] +
        \left[\hat{r}_{21}(z_2,z_1),I \otimes l(z_2)\right]
    \label{eq:poisson}
\end{equation}
where
\[
    \hat{r}_{12}(z_1,z_2) = {z_2 \over z_1-z_2}t + 
                            z_2 \sum_{jk} E_{jk}U^{N+1-j} \otimes E_{k1}
\]

The equation (\ref{eq:poisson}) specifies the Poisson structure on the matrix
elements which we are going to use henceforth.

Note that we could have started directly from this expression of the Poisson
brackets, however obtaining it from the considerations used above gives some
additional insight into why do we have to use the $r$-matrix $\hat{r}_{12}$
here.

%% file: separation.tex
\newcommand{\Dm}{{\Delta}m(z_1,z_2)}
\newcommand{\PhiT}{\transpose{\Phi}}
\newcommand{\ahat}{\hat{a}}
\newcommand{\dhat}{\hat{d}}
\newcommand{\Xhat}{\hat{X}}
\newcommand{\Yhat}{\hat{Y}}
\newcommand{\Zhat}{\hat{Z}}
\newcommand{\Ismall}{\tilde{I}}
\newcommand{\Ksmall}{\tilde{K}}
\newcommand{\Usmall}{\tilde{U}}

\section{Separation of Variables}
\label{sec:separation}

In this section we are going to construct the separated variables on the
Jacobian from the matrix $l(z)$. This is the first step of our plan to
establish an isomorphism between the (affine) Jacobian and the space of these
matrices, although it also provides a generalization of the construction of
the separated variables originally proposed in \cite{Sklyanin}.

\subsection{The Baker--Akhiezer function}

Consider the function $\psi$ which is the eigenvector of $l(z)$ for the
eigenvalue $-w$:
\begin{equation}
    \left(l(z) + w\right) \psi(w, z) = 0
    \label{eq:bafunc}
\end{equation}

Any of the components of the vector $\psi(w, z)$ is a Baker--Akhiezer function.
The poles of $\psi_i(w, z)$ are fixed (they are the points of the dynamic
divisor describing the time evolution of the system at the moment $t = 0$) and
the zeroes of them give the separated coordinates as we'll see below.

Let us choose the first component and let $(w_i, z_i)$ be the $g$ zeroes of
$\psi_1(w, z)$. Then write $l(z)$ as:
\[
    l(z) = \left(
            \begin{tabular}{c|c}
                \raisebox{0pt}[12pt][10pt]{$l_{11}(z)$}
                    & \makebox[3\width]{$b(z)$} \\
                \multispan2 \hrulefill \\
                $*$ \\
                $\vdots$ & $d(z)$ \\
                $*$
                \end{tabular}
           \right)
\]
with the notations inspired by $N=2$ case where the matrix was written as
\[
    l(z) = \left(
            \begin{array}{cc}
                a(z) & b(z) \\
                c(z) & d(z)
            \end{array}
           \right)
\]

Let us also split $\psi$ likewise:
\[
    \psi(w, z) = \left(
                \begin{array}{c}
                    \psi_1(w, z) \\
                    \psi_2(w, z) \\
                    \vdots \\
                    1
                \end{array}
              \right)
            = \left(
                \begin{array}{c}
                    \psi_1(w, z) \\
                    \phi(w, z)
                \end{array}
              \right)
\]

Then for any $z_i$ decomposing the equation (\ref{eq:bafunc}) defining
$\psi$ gives
\[
    \left\{
        \begin{array}{lcl}
            b(z_i) \phi(w_i, z_i) & = & 0 \\
            d(z_i) \phi(w_i, z_i) & = & -w_i \phi(z_i)
        \end{array}
    \right.
\]
Or, combining them:
\begin{equation}
    b(z_i) d^k(z_i) \phi(w_i, z_i) = 0 \qquad \forall k \geq 0
    \label{eq:lznull}
\end{equation}

This leads us to the idea to consider the matrix
\[
    Z(z) = \left(
            \begin{array}{c}
                b(z)            \\
                b(z) d(z)       \\
                \vdots          \\
                b(z) d^{N-2}(z)
            \end{array}
           \right)
\]

(\ref{eq:lznull}) implies, of course, that $Z(z_i) \phi(w_i, z_i) = 0$ for any
$i$ and hence, as $\phi \neq 0$ (its last component is $1$) we have
\[
    \det Z(z_i) = 0 \qquad \forall i = 1, \ldots, g
\]

Remembering the specific form of degrees of the matrix elements from
(\ref{eq:ldegree}) we can see that the maximal degree of $z$ in $\det Z(z)$
comes from the product containing $l_{1N}$
and the elements from the second main diagonal in the matrix $l(z)$. This
degree is equal to $\sum_{k=1}^{N-2}(kn-1) = g$ and so the determinant
$\det Z(z)$ is a polynomial in $z$ of degree $g$ which has $g$ roots
$z_i$ which proves that it is equal (up to a constant factor) to
$\prod_i (z - z_i)$.

This allows us to now define $z_i$ as the roots of the polynomial $\det Z(z)$
which is constructed from the matrix elements alone (and without any further
reference to the Baker--Akhiezer function). Again, we could have started by
defining them in this way as the subsequent calculations only use the properties
of the matrix $Z$ but making the link with the Baker--Akhiezer functions makes
it more clear why do we need to define $z_i$ in this way.

\subsection{Determination of $w_i$}

Next we are going to show that $w_i$ are unambiguously determined by the
matrix elements as well.

For this, let us introduce the $N \times (N-1)$ matrix
\[
     \left(
        \begin{array}{c}
            b(z) \\
            d(z) + w
        \end{array}
     \right)
\]
Considering its action on the vector $\phi(w_i, z_i)$ we see that the $N$
determinants obtained by removing the $k$-th row from this matrix for $k = 1,
\ldots, N$ are null. Thus, for any $(N-1) \times N$ matrix $A$ we have
\begin{equation}
    \det \left( A \left(
                    \begin{array}{c}
                        b(z_i) \\
                        d(z_i) + w_i
                    \end{array}
                  \right) \right) = 0
    \label{eq:AnyDetIsNull}
\end{equation}

Let us apply this with
\[
     A = \left(
            \begin{array}{cc}
                1       & 0 \ldots 0            \\
                0       & b(z_i)                \\
                0       & b(z_i) d(z_i)         \\
                \vdots  & \vdots                \\
                0       & b(z_i) d^{N-4}(z_i)   \\
                0       & \xi                             
            \end{array}
         \right)
\]
where $\xi$ is an arbitrary row vector.

Then (\ref{eq:AnyDetIsNull}) yields, after eliminating the other terms by line
subtraction,
\begin{equation}
    \det \left(
            \begin{array}{c}
                b(z_i) \\
                b(z_i) d(z_i) \\
                \vdots \\
                b(z_i) d^{N-3}(z_i) \\
                \xi (d(z_i) + w_i)
            \end{array}
         \right) = 0
    \label{eq:charw}
\end{equation}

At this point we need to make an additional genericity hypothesis: let us
suppose that $z_i$ is a simple root of $\det Z(z)$. This ensures that the
matrix
\[
    \left(
            \begin{array}{c}
                b(z_i) \\
                b(z_i) d(z_i) \\
                \vdots \\
                b(z_i) d^{N-3}(z_i) \\
            \end{array}
    \right)
\]
is of rank $N-2$. Indeed, if this were not the case we would have had
\[
    \sum_{k=1}^{N-2} b(z_i) d^{k-1}(z_i) = 0
\]
and hence
\[
    \sum_{k=1}^{N-2} b(z) d^{k-1}(z) = \Ob\left(z - z_i\right)
\]
and also
\[
    \sum_{k=1}^{N-2} b(z) d^{k}(z) = \Ob\left(z - z_i\right)
\]
but the last two equations imply
\[
    \det Z(z) = \Ob\left((z - z_i)^2\right)
\]
which contradicts our hypothesis.

Thus this hypothesis ensures that there exists a $\xi$ for which $w_i$ can be
defined by
\begin{equation}
    w_i = - {
              \det \left(
                           \begin{array}{c}
                               b(z_i) \\
                               b(z_i) d(z_i) \\
                               \vdots \\
                               b(z_i) d^{N-3}(z_i) \\
                               \xi d(z_i)
                           \end{array}
                   \right)
              \over
              \det \left(
                           \begin{array}{c}
                               b(z_i) \\
                               b(z_i) d(z_i) \\
                               \vdots \\
                               b(z_i) d^{N-3}(z_i) \\
                               \xi
                           \end{array}
                   \right)
            }
    \label{eq:defw}
\end{equation}
as, clearly, then (\ref{eq:charw}) is satisfied.

So we have achieved our goal of constructing $g$ points $(w_i, z_i)$ from the
given matrix elements.

Note that although $w_i$ can still be uniquely constructed even in presence of
double roots of $\det Z(z)$, we would have other, more serious, problems with
reconstructing the matrix elements in the next section without the genericity
hypothesis and so we will assume it from now on.

\subsection{The Poisson brackets}

So far we have seen that, starting from the matrix elements, we may define a
point $(w_i, z_i)$ on the jacobian. But it still remains to show that $w$ and
$z$ defined in this way satisfy the canonical commutation relations and so are
really separated variables in our problem.

These commutation relations have been first calculated in \cite{Dubrovin} but,
as mentioned in the introduction, we propose a more algebraic way of
computing them which allows a relatively straightforward generalization to the
quantum case as seen in \cite{qSmirnov}.

So in this section we are going to carry out the calculations proving that we
indeed have the expected commutation relations for the variables we have just
constructed. To simplify them, first let us note that we can define $Z(z)$ in
terms of the matrix $m(z)$ instead of $l(z)$ as conjugation by $s$ doesn't
change the brackets of the determinants but the calculations which follow
become much simpler because of a simpler form of the brackets of $m(z)$
itself.

Thus in the rest of this section we are going to redefine $b(z)$ and $d(z)$ as
the corresponding parts of the matrix $m(z)$ and not $l(z)$, in other words
we have
\[
    m(z) = \left(
            \begin{array}{cc}
                m_{11}(z)   & b(z) \\
                \vdots      & d(z)
            \end{array}
           \right)
\]
where $b(z)$ a covector of size $N-1$ and $d(z)$ is a square matrix of order
$N-1$.

To abbreviate the notations, we will omit the argument $z_1$ and use primes
for the functions of $z_2$. We will also use the short hand notations: \[
    \left\{
        \begin{array}{lclcl}
            x_1 & \equiv & x(z_1) \otimes I & = & x \otimes I \\
            x_2 & \equiv & I \otimes x(z_2) & = & I \otimes x'
        \end{array}
    \right.
\]
and
\[
    \tpoisson{x(z_1)}{x(z_2)}_{ij, kl} =
        \{x_1, x_2\}_{ij,kl} = \{x_{ik}, x'_{jl}\}
\]
for any $x$.

Finally, we introduce a special notation for the rows of the matrix $Z(z)$:
\[
    a^{(k)}(z) \equiv b(z) d^{k-1}(z)
\]

\subsubsection{The brackets $\{z_i, z_j\}$}

Our first goal is to show that
\[
    \left\{\det Z, \det Z'\right\} = 0
\]

For this first let us use the identity
\begin{eqnarray}
\ifnum\IncludeDetails=1
    \lefteqn{\left\{\det Z, \det Z'\right\} = } \nonumber \\
    & = & \frac{\partial \det Z}{\partial Z_{ij}}
          \frac{\partial \det Z'}{\partial Z'_{kl}}
          \left\{Z_{ij}, \det Z'_{kl}\right\} \nonumber \\
    & = & \det Z \det Z' (Z^{-1})_{ji} (Z'^{-1})_{lk}
          \left\{Z_{ij}, \det Z'_{kl}\right\} \nonumber \\
    & = & \det Z \det Z' \left(
            \left(Z^{-1} \otimes Z'^{-1}\right) \tpoisson{Z}{Z'}
          \right)_{jl,jl} \nonumber \\
\else
    \left\{\det Z, \det Z'\right\}
\fi
    & = & \det Z \det Z' \left(\tr \otimes \tr\right) \left(
            Z_1^{-1} Z_2^{-1} \{ Z_1, Z_2 \}
          \right)
    \label{eq:detbrack}
\end{eqnarray}
which shows that we need to calculate $\{Z_1, Z_2\}$ and for this, in turn, we
need to know $\{b_1, b_2\}$, $\{b_1, d_2\}$ and $\{d_1, d_2\}$. To compute
them we start with the definition (\ref{eq:brackets})
\begin{equation}
    \{m_1, m_2\} = \left[r(z_1,z_2), \Dm \right]
    \label{eq:mbrack}
\end{equation}
where
\[
    \Dm \equiv m_1 + m_2 = m \otimes I + I \otimes m'
\]
and split this relation according to the decomposition of the matrix $m(z)$:
let us introduce the matrix $\rho$ which is the same as $r(z_1,z_2)$ but of
order $N-1$ (to be precise, $\rho_{ij,kl} = r_{i+1\,j+1,k+1\,l+1}$). Then we
immediately get from (\ref{eq:mbrack}):
\[
    \{d_1, d_2\} = \left[\rho(z_1,z_2), d_1 + d_2\right]
\]

As $\Dm_{ij,kl} = 0$ unless $i = k$ or $j = l$ and $r_{ij,kl} \neq 0$ only if
$i = l$ and $j = k$ we see that (\ref{eq:mbrack}) also gives
\ifnum\IncludeDetails=1
\begin{eqnarray*}
    \{ m_{1,i+1}, m_{1,j+1}' \}
        & = & \left[r(z_1,z_2), \Dm_{11,i+1\,j+1}\right] \\
        & = & r(z_1,z_2)_{11,kl} \Dm_{kl,i+1\,j+1} - \\
        &   & \Dm_{11,kl} r(z_1,z_2)_{kl,i+1\,j+1} \\
        & = & \Dm_{11,i+1\,j+1} r(z_1,z_2)_{11,11} - \\
        &   & \Dm_{11,j+1\,i+1} r(z_1,z_2)_{j+1\,i+1,i+1\,j+1} \\
        & = & 0
\end{eqnarray*}
and so
\fi
\[
    \{b_1, b_2\} = 0
\]

And using the explicit form of $r$ we obtain
\[
    \left\{ b_1, d_2 \right\} = {z_2 \over z_1-z_2} \Phi b(z_2) - b_1 \rho
\]
The object $\Phi$ appearing here is a notational convenience and is formally
defined as
\[
    \Phi = \sum_{i=1}^N e_i \otimes f_i
\]
with $e_i$ and $f_i$ being a row and column matrices respectively with $1$ at
$i$-th position and $0$ elsewhere. $\Phi$ transposes the matrix dimension in
the tensor product, i.e. when it is applied to $b_2$ which is a matrix in the
first component and a row in the second one it yields a tensor product of a
row in the first component and a matrix in the second one defined by
\[
    (\Phi b_2)_{i,jk} = \delta_{ij} b'_k
\]
We also define $\PhiT$ which results in a tensor product of a matrix and a row
and which acts on $b_1$ as
\[
    (\PhiT b_1)_{i, jk} = \delta_{ik} b_j
\]

Our next goal is to calculate $\{b_1 d_1^k, b_2 d_2^l\}$ for which we need
\begin{eqnarray*}
    \{ d_1^k, d_2^l \}
        & = & \left( \sum_{p=1}^l d_2^{p-1} \rho d_2^{l-p} \right) d_1^l +
              \left( \sum_{q=1}^k d_1^{q-1} \rho d_1^{k-q} \right) d_2^l - \\
        & - & d_1^k \left( \sum_{p=1}^l d_2^{p-1} \rho d_2^{l-p} \right) -
              d_2^l \left( \sum_{q=1}^k d_1^{q-1} \rho d_1^{k-q} \right)
\end{eqnarray*}
(as the other terms simply cancel pairwise) and also
\begin{eqnarray*}
    \{ b_1, d_2^l \} & = & -b_1 \sum_{p=1}^l d_2^{p-1} \rho d_2^{l-p} +
            \sum_{p=1}^l d_2^{p-1} \frac{z_2}{z_1-z_2} \Phi b_2 d_2^{l-p} \\
    \{ d_1^k, b_2 \} & = & -b_2 \sum_{q=1}^k d_1^{q-1} \rho d_1^{k-q} +
            \sum_{q=1}^k d_1^{q-1} \frac{z_1}{z_1-z_2} \PhiT b_1 d_1^{k-q}
\end{eqnarray*}

Let us define some more abbreviations:
\begin{eqnarray*}
    \left\{
        \begin{array}{lcl}
            \dhat^{(k)}_{12} & \equiv & \sum_{q=1}^{k-1} d^{q-1}_1 \rho d^{k-q-1}_1 \\
            \dhat^{(l)}_{21} & \equiv & \sum_{p=1}^{l-1} d^{p-1}_2 \rho d^{l-p-1}_2
        \end{array}
    \right.
\end{eqnarray*}
and
\begin{eqnarray*}
    \left\{
        \begin{array}{lcl}
            \ahat^{(k)}_{12} & \equiv & b_1 \dhat^{(k)}_{12} \\
            \ahat^{(l)}_{21} & \equiv & b_2 \dhat^{(l)}_{21} \\
        \end{array}
    \right.
\end{eqnarray*}

Putting all the above together and using the new notations we obtain:
\begin{eqnarray}
    \lefteqn{\left\{ b_1 d_1^{k-1}, b_2 d_2^{l-1} \right\} = } \nonumber \\
    & = & b_2 \left\{ b_1, d_2^{l-1} \right\} d_1^{k-1} +
            b_1 \left\{ d_1^{k-1}, b_2 \right\} d_2^{l-1} +
            b_1 b_2 \left\{ d_1^{k-1}, d_2^{l-1} \right\} \nonumber \\
    & = & -a^{(l)}_2 \ahat^{(k)}_{12} -a^{(k)}_1 \ahat^{(l)}_{21} + \nonumber \\
    & + & \sum_{p=1}^{l-1} a_2^{(p)}
            \frac{z_2}{z_1-z_2} \Phi a_2^{(l-p)} d_1^{k-1} +
          \sum_{q=1}^{k-1} a_1^{(q)}
            \frac{z_1}{z_2-z_2} \PhiT a_1^{(k-q)} d_2^{l-1}
    \label{eq:Zrowbrack}
\end{eqnarray}

Now we have all ingredients to calculate the brackets of $Z(z)$ itself: for
this we just rewrite (\ref{eq:Zrowbrack}) in a matrix form:
\begin{eqnarray}
    \{Z_1, Z_2\} & = & - Z_1 \Zhat_{21} - Z_2 \Zhat_{12} + \nonumber \\
    & + & \frac{1}{z_1-z_2} \sum_{r,s}
            \left( z_2 Z'_{rs}
                \left(D_s \otimes U^r Z'\right)
            \right. +
            \left.
                   z_1 Z_{rs} \left(U^r Z \otimes D'_s\right)
            \right)
    \label{eq:ZZbrack}
\end{eqnarray}
where $\Zhat_{21}$ is a matrix whose rows (relative to the second
tensor product component) are $\ahat^{(k)}_{21}$:
\[
    \Zhat_{21}(z_1, z_2) = \sum_{k=1}^{N-1}
                                \left(I \otimes f_k\right) \ahat^{(k)}_{21}
\]
and $\Zhat_{12}(z_1, z_2)$ is defined in the same way using $\ahat^{(l)}_{12}$:
\[
    \Zhat_{12}(z_1, z_2) = \sum_{l=1}^{N-1}
                                \left(f_l \otimes I\right) \ahat^{(l)}_{12}
\]

Finally, $D_r$ is just the matrix tailored to allow us to pick the $r$-th
element of $d$ in the matrix form:
\begin{eqnarray*}
    D_r(z) = \left(
                \begin{array}{c}
                    e_r         \\
                    e_r d(z)    \\
                    \vdots      \\
                    e_r d^{N-2}(z)
                \end{array}
             \right)
\end{eqnarray*}
and the usage of $U$ (defined by (\ref{eq:Umatrix})) lets us to sum over all
values of $r$ instead of just the ones in range $1 \leq r \leq i$ (for matrix
element $ij,kl$).

We can now inject (\ref{eq:ZZbrack}) into (\ref{eq:detbrack}). Let us consider
the four terms resulting of (\ref{eq:ZZbrack}) independently, starting with
the first one (containing $\Zhat_{21}$) which results in:
\[
    \det Z \det Z' \left(\tr \otimes \tr\right) \left(
        Z_2^{-1} \Zhat_{21}(z_2, z_1)
    \right)
\]
To calculate $\tr \otimes I$ note that only $\rho$ contributes and hence,
using its explicit form:
\[
    \left(\tr \otimes I\right) \rho(z_1, z_2) = \half \frac{z_1+z_2}{z_1-z_2} I
\]
so, leaving the factors aside, we have only
\[
    \left(I \otimes \tr\right) Z_2^{-1} \Zhat_{21}
\]
to calculate. And to do this, we simply notice that, after eliminating $\rho$
from $\ahat^{(k)}_{21}$ we are left with $(k - 1) b_2 d^{(k-1)}$ or, in other words,
the $(k-1)$-th line of $Z_2$ times $k - 1$ and so this trace reduces to
\[
    \tr \left(K Z_2 Z_2^{-1}\right)
\]
with
\begin{equation}
    K = \left(
            \begin{array}{ccccc}
            0 & 0 & 0 & \ldots & 0 \\
            1 & 0 & 0 &        & 0 \\
            0 & 2 & 0 &        & 0 \\
            \vdots &   & \ddots & \ddots & \vdots \\
            0 & \ldots & \ldots & N-2 & 0
            \end{array}
        \right)
    \label{eq:Kmatrix}
\end{equation}
which of course means, as $\tr K = 0$, that the contribution from the first
term is $0$. The contribution of the second term vanishes in exactly the same
way.

Now consider the third term which contains a sum of
\[
    \left(Z^{-1} \otimes Z'^{-1}\right) Z'_{rs} \left(D_s \otimes U^r Z'\right)
\]
The only contribution to the trace over the second space, after $Z'$ cancel,
is $\tr U^r$ which yields a $0$ as a factor and so this term vanishes as well
- just as the fourth one does for the symmetrical reasons.

So we have finally established that
\[
    \left\{\det Z, \det Z'\right\} = 0
\]
which proves that $z_i$ defined as the roots of $\det Z(z)$ also satisfy
\[
    \left\{z_i, z_j\right\} = 0 \quad \forall i,\,j
\]

\subsubsection{The brackets $\{z_i, w_j\}$}

The next step is to calculate the (only non trivial) brackets of $z$ and $w$.
Let us define $B(z)$ and $A(z)$ by
\[
    \left\{
        \begin{array}{lcll}
            B(z_i) & = & 0 & \quad \quad \hbox{i.e. } B(z) \equiv \det Z(z) \\ 
            w_i & = & -A(z_i)
        \end{array}
    \right.
\]
As we have seen above, $B(z)$ is a polynomial in the matrix elements and
$A(z)$ is a rational fraction in them (defined by (\ref{eq:defw})). We use
these notations just to emphasize that these $B$ and $A$ are the
generalizations of the functions from \cite{Sklyanin}.

Then we have
\[
    0 = \left\{ B(z_i), w_j \right\}
      = \left\{ B(z), w_j \right\} \\|_{z=z_i} + B'(z_i) \left\{ z_i, w_j \right\}
\]
and so
\begin{equation}
    \{z_i, w_j\} = - \frac{\{B(z), A(z_j)\}\\|_{z=z_i}}{B'(z_i)}
    \label{eq:zwbrack}
\end{equation}
which means that our task is to calculate $\{B(z), A(z')\}$.

For this we will reuse the results of the previous subsection: note
that $A(z)$ can be written as
\[
    A(z) = { \det X(z) \over \det Y(z) }
\]
with
\[
    X \equiv \left(
                \begin{array}{c}
                    a^{(1)}     \\
                    \vdots      \\
                    a^{(N-2)}   \\
                    \xi d
                \end{array}
             \right)
    \quad \hbox{ and } \quad
    Y \equiv \left(
                \begin{array}{c}
                    a^{(1)}     \\
                    \vdots      \\
                    a^{(N-2)}   \\
                    \xi
                \end{array}
             \right)
\]
and so to calculate the brackets in question we have to calculate the brackets
of $Z$ with $X$ and $Y$ which is done similarly to the calculation above with
the only differences being in the last row.

Accordingly we are going to proceed in the same order as above so as to obtain
a form similar to (\ref{eq:ZZbrack}). Thus we first define the matrices
corresponding to $\Zhat$:
\[
    \Xhat_{21} \equiv \left(
                        \begin{array}{c}
                            \ahat^{(1)}_{21}    \\
                            \vdots              \\
                            \ahat^{(N-2)}_{21}  \\
                            \xi_2 \rho
                        \end{array}
                      \right)
    \quad \hbox{ and } \quad
    \Yhat_{21} \equiv \left(
                        \begin{array}{c}
                            \ahat^{(1)}_{21}    \\
                            \vdots              \\
                            \ahat^{(N-2)}_{21}  \\
                            0
                        \end{array}
                      \right)
\]
Note that the rows here are relative to the second tensor product component,
$\ahat_{21}$ already being matrices in the first space. And $\Xhat_{12}$ and
$\Yhat_{12}$ are defined in the same way using $\ahat_{12}$ and $\xi_1$ for
the last row (relative to the first space) of $\Xhat_{12}$.

Let us start with the easier computation, that of $\{Z_1, Y_2\}$: in this case
we already know the result as it will be the same as (\ref{eq:ZZbrack})
for all rows (relative to the second space) except for the last one for which
it is simply null. So we can write
\begin{eqnarray}
    \{ Z_1, Y_2 \}
        & = & - Z_1 \Yhat_{21} - Y_2 \Zhat_{12} + \nonumber \\
        & + & \frac{1}{z_1-z_2} \sum_{r,s}
                \left( z_2 Y'_{rs}
                    \left(D_s \otimes \Usmall^r Y'\right)
                \right. +
                \left.
                       z_1 Z_{rs} \left(U^r Z \otimes \Ismall D'_s\right)
                \right) + \nonumber \\
        & + & (I \otimes f_{N-1}) \xi_2 \Zhat_{12}
    \label{eq:ZYbrack}
\end{eqnarray}
with $\Ismall$ and $\Usmall$ being, respectively, the identity matrix and $U$
without the last row:
\[
    \Ismall = \left(
                \begin{array}{cccc}
                    1 \\
                        & \ddots \\
                        &           & 1 \\
                        &           &   & 0
                \end{array}
              \right)
    \quad \quad
    \Usmall = \left(
                \begin{array}{ccccc}
                    0 & 0 & 0 & \ldots & 0 \\
                    1 & 0 & 0 &        & 0 \\
                    0 & 1 & 0 &        & 0 \\
                    \vdots &   & \ddots & \ddots & \vdots \\
                    0 & \ldots & \ldots & 0 & 0
                \end{array}
              \right)
\]
Note that $U^r Z = U^r Y = U^r X$ for any $r$ anyhow and we only use $U^r Y$
above for cosmetic reasons.

Using again the identity (\ref{eq:detbrack}) we get only the contribution from
the last term of (\ref{eq:ZYbrack}):
\begin{equation}
    \left\{ \det Z_1, \det Y_2 \right\} =
        \det Z_1 \det Y_2 {\tr}_2 \left(1 \otimes f_{N-1}\right) \eta_2 Y_2^{-1}
    \label{eq:ZYdetbrack}
\end{equation}
with
\[
    \eta_2 \equiv \xi_2 {\tr}_1 \left( Z_1^{-1} \Zhat_{12} \right)
\]
as all the other terms vanish inside $\tr \otimes \tr$ for the same reasons as
in the previous subsection.

Now let us turn to the brackets of $Z$ and $X$. Here we again have, of course,
the same expression (\ref{eq:ZZbrack}) for all rows except the last one where
there must be (leaving aside $\xi_2$)
\begin{equation}
    \left\{ a^{(k)}_1, d_2 \right\} =
        \frac{z_2}{z_1-z_2} \Phi b_2 d_1^{k-1} -
        a^{(k)}_1 \rho +
        \left[ \ahat^{(k)}_{12}, d_2 \right]
    \label{eq:adbrack}
\end{equation}

This leads us to the following result:
\begin{eqnarray}
    \{ Z_1, X_2 \}
        & = & - Z_1 \Xhat_{21} - X_2 \Zhat_{12} + \nonumber \\
        & + & \frac{1}{z_1-z_2} \sum_{s}
                z_2
                    D_s \otimes \left(
                                        \left(
                                            \sum_r X'_{rs} \Usmall^r
                                        \right) + \xi_s J
                                \right) X'
                + \nonumber \\
        & + & \frac{1}{z_1-z_2} \sum_{r, s}
                z_1 Z_{rs} \left(U^r Z \otimes \Ismall D'_s\right)
              + \nonumber \\
        & + & (I \otimes f_{N-1}) \xi_2 \Zhat_{12} d_2
    \label{eq:ZXbrack}
\end{eqnarray}
where we again needed a new notation - this time for the matrix
\[
    J \equiv \left(
                \begin{array}{ccc}
                    0       & \cdots    & 0             \\
                    \vdots  &           & \vdots        \\
                    1       & \cdots    & 0
                \end{array}
             \right)
\]

The term containing $J$ cancels with the first term of (\ref{eq:adbrack})
while the combination of the remaining terms of the LHS and $Z_1 \Xhat_{21}$
gives exactly the last term of (\ref{eq:ZXbrack}).

We now want to calculate $\{ \det Z_1, \det X_2 \}$ using the identity
(\ref{eq:detbrack}) again. We have two yet unknown contributions to it which
we need to compute now. The first one is coming from $\Xhat_{21}$ and the
second one from the last term of (\ref{eq:ZXbrack}). For the latter, we are
not going to compute it at all but just rewrite it using $\eta_2$ defined
above, then its contribution to the brackets is
\begin{equation}
    \det Z_1 \det X_2 {\tr}_2 \left(I \otimes f_{N-1}\right) \eta_2 d_2 X^{-1}_2
    \label{eq:detZXbrack}
\end{equation}

To compute the former, first note, referring to the calculation done for
${\tr}_1 \Zhat_{21}$ above that
\[
    {\tr}_1 \Xhat_{21} =
        \half \frac{z_1+z_2}{z_1-z_2}
            \left(
                \Ksmall X_1 + \left(I \otimes f_{N-1}\right) \xi_2
            \right)
\]
with $\Ksmall$ being the matrix $K$ defined by (\ref{eq:Kmatrix}) without the
last row:
\[
    \Ksmall = \left(
                \begin{array}{ccccc}
                    0 & 0 & 0 & \ldots & 0 \\
                    1 & 0 & 0 &        & 0 \\
                    0 & 2 & 0 &        & 0 \\
                    \vdots & & & \ddots & \vdots \\
                    0 & \ldots & N-2 & 0 & 0 \\
                    0 & \ldots & \ldots & 0 & 0
                \end{array}
              \right)
\]
As above, the first term doesn't contribute as $\tr \Ksmall = 0$ and the
second one yields
\[
    {\tr}_2 \left( {\tr}_1 \Xhat_{21} \right) X^{-1}_2 =
        \half \frac{z_1+z_2}{z_1-z_2} \tr \left( f_{N-1} \xi (X')^{-1} \right)
\]
To calculate the latter trace we decompose $\xi$ over the basis of the rows of
$X$
\[
    \xi = \sum_{k=1}^{N-2} \lambda_k a'^{(k)} + \lambda \xi d'
\]
and then we have
\begin{eqnarray*}
    \tr \left( f_{N-1} \xi (X')^{-1} \right)
        & = & \sum_{k=1}^{N-2} \lambda_k (X' X'^{-1})_{k,N-1} +
              \lambda (X' X'^{-1})_{N-1, N-1} \\
        & = & \lambda \\
        & = & \frac{a'^{(1)} \wedge \cdots \wedge a'^{(N-2)} \wedge \xi}
                   {a'^{(1)} \wedge \cdots \wedge a'^{(N-2)} \wedge \xi d'} \\
        & = & \frac{\det Y'}{\det X'}
\end{eqnarray*}

So combining all non vanishing contributions we obtain
\begin{eqnarray*}
    \left\{ \det Z_1, \det X_2 \right\}
        & = & - \half \frac{z_1+z_2}{z_1-z_2} \det Z_1 \det Y_2 + \\
        & + & \det Z_1 \det X_2
                {\tr}_2 \left(1 \otimes f_{N-1}\right) \eta_2 d_2 X_2^{-1}
    \label{eq:ZXdetbrack}
\end{eqnarray*}

Now putting together (\ref{eq:ZYdetbrack}) and (\ref{eq:ZXdetbrack}) we have
\begin{eqnarray*}
    \left\{ \det Z_1, \frac{\det X_2}{\det Y_2} \right\}
        & = & - \half \frac{z_1+z_2}{z_1-z_2} \det Z_1 + \\
        & + & \left[
                {\tr}_2 \left(1 \otimes f_{N-1}\right)
                    \left(
                        \eta_2 d_2 X_2^{-1} - \eta_2 Y_2^{-1}
                    \right)
              \right] \det A_1 \frac{\det X_2}{\det Y_2}
\end{eqnarray*}
which can be further simplified if we now decompose $\eta$ in the basis of
rows of $Y'$:
\[
    \eta = \sum_{k=1}^{N-2} \mu_k a'^{(k)} + \mu \xi
\]
Then:
\[
    \eta d' = \sum_{k=1}^{N-3} \mu_k a'^{(k + 1)} +
                \mu_{N-2} a'^{(N-1)} + \mu \xi d'
\]
and
\begin{eqnarray*}
    \lefteqn{
                {\tr}_2 \left(1 \otimes f_{N-1}\right)
                \left(
                    \eta_2 d_2 X_2^{-1} - \eta_2 Y_2^{-1}
                \right) =
            } \\
        & = & \sum_{k=1}^{N-2}
                \left(
                    \mu_k (X' X'^{-1})_{k+1, N-1} -
                    \mu_k (Y' Y'^{-1})_{k, N-1}
                \right) + \\
        & + & \mu_{N-1} \left(
                            (X' X'^{-1})_{N-1,N-1} -
                            (Y' Y'^{-1})_{N-1,N-1}
                        \right) + \\
        & + & \mu_{N-2} {\tr}_2 \left(I \otimes f_{N-1}\right)
                a'^{(N-1)} X'^{-1} \\
        & = & \mu_{N-2} (Z' X'^{-1})_{N-1, N-1} \\
        & = & \mu_{N-2} \frac{\det Z_2}{\det X_2}
\end{eqnarray*}

Taking account of this we finally get
\begin{equation}
    \left\{ B(z_1), A(z_2) \right\} =
        - \half \frac{z_1+z_2}{z_1-z_2} B(z_1) +
        \left( \mu_{N-2} \frac{\det Z(z_1)}{\det Y(z_2)} \right) B(z_2)
    \label{eq:ZXYdetbrack}
\end{equation}
Injecting this result into (\ref{eq:zwbrack}) we see that the brackets vanish
for $z_1 = z_i$, $z_2 = z_j$ as they are both roots of $B(z)$ unless $i = j$.
In this case simply taking the limit $z \to z_j$ shows that we finally have
\[
    \left\{ z_i, w_j \right\} = \delta_{ij} z_i
\]

\subsubsection{The brackets $\{w_i, w_j\}$}

The final calculation we have to do is to show that
\[
    \{ w_i, w_j \} = 0
\]
or, in another words, that
\begin{equation}
    \left\{ A(z_i), A(z_j) \right\} = 0 \qquad \forall i, j
    \label{eq:AAbrack}
\end{equation}

To calculate the brackets of $A(z)$ we need, of course, all the brackets of
$X$ and $Y$ between themselves and with each other. And for this we are again
going to proceed as above by reusing the results obtained for $\{ Z_1, Z_2 \}$
for all the rows but the last one.

Let us start by calculating the brackets of $X$. We are going to have the
following expression in the last row:
\[
    \left\{ \xi_1 d_1, \xi_2 d_2 \right\} =
        \xi_1 \xi_2 \left[ \rho, d_1 + d_2 \right]
\]
and so, in a way similar to (\ref{eq:ZXbrack}), we find
\begin{eqnarray}
    \left\{ X_1, X_2 \right\}
        & = & - X_1 \Xhat_{21} - X_2 \Xhat_{12} + \nonumber \\
        & + & \frac{1}{z_1-z_2} \sum_{s}
                z_2
                    \Ismall D_s \otimes
                    \left(
                        \left(
                            \sum_r X'_{rs} \Usmall^r
                        \right) + \xi_s J
                    \right) X'
                + \nonumber \\
        & + & \frac{1}{z_1-z_2} \sum_{s}
                z_1
                    \left(
                        \left(
                            \sum_r X_{rs} \Usmall^r
                        \right) + \xi_s J
                    \right) X
                    \otimes \Ismall D'_s
                + \nonumber \\
        & + & (I \otimes f_{N-1}) \xi_2 \Xhat_{12} d_2 +
              (f_{N-1} \otimes I) \xi_1 \Xhat_{21} d_1
    \label{eq:XXbrack}
\end{eqnarray}

By repeating the steps which resulted in (\ref{eq:detZXbrack}), we have:
\begin{eqnarray}
    \frac{ \left\{ \det X_1, \det X_2 \right\} }{ \det X_1 \det X_2 }
        & = & - \half \frac{z_1 + z_2}{z_1 - z_2}
                    \left(
                        \frac{\det Y_2}{\det X_2} + \frac{\det Y_1}{\det X_1}
                    \right) \nonumber \\
        & + & {\tr}_2
              \left(
                (I \otimes f_{N-1}) \xi_2 d_2 X^{-1}_2
              \right) + \nonumber \\
        & + & {\tr}_1 \left( (f_{N-1} \otimes I) \xi_1 d_1 X^{-1}_1 \right)
    \label{eq:detXXbrack}
\end{eqnarray}
with
\[
    \left\{
        \begin{array}{lcl}
            \zeta_1 & \equiv & {\tr}_1 \xi_2 ( X^{-1}_1 \Xhat_{12} ) \\
            \zeta_2 & \equiv & {\tr}_2 \xi_1 ( X^{-1}_2 \Xhat_{21} )
        \end{array}
    \right.
\]

In exactly the same way we also obtain
\begin{eqnarray}
    \frac{ \left\{ \det X_1, \det Y_2 \right\} }{ \det X_1 \det Y_2 }
        & = & - \half \frac{z_1 + z_2}{z_1 - z_2} \frac{\det Y_1}{\det X_1} + \nonumber \\
        & + & {\tr}_2 \left( (I \otimes f_{N-1}) \xi_2 Y^{-1}_2 \right) + \nonumber \\
        & + & {\tr}_1 \left( (f_{N-1} \otimes I) \kappa_1 d_1 X^{-1}_1 \right)
    \label{eq:detXYbrack}
\end{eqnarray}
with
\[
    \kappa_1 \equiv {\tr}_2 \left( \xi_1 \Yhat_{21} Y^{-1}_2 \right)
\]   
and a symmetrical expression for $\{ \det Y_1, \det X_2 \}$.

Finally, we also find
\begin{eqnarray}
    \frac{ \left\{ \det Y_1, \det Y_2 \right\} }{ \det Y_1 \det Y_2 }
        & = & {\tr}_2 \left( (I \otimes f_{N-1}) \kappa_2 Y^{-1}_2 \right) + \nonumber \\
        & + & {\tr}_1 \left( (f_{N-1} \otimes I) \kappa_1 Y^{-1}_1 \right)
    \label{eq:detYYbrack}
\end{eqnarray}

Combining (\ref{eq:detXXbrack}), (\ref{eq:detXYbrack}) and
(\ref{eq:detYYbrack}) together we may calculate
\begin{eqnarray*}
    \frac{ \left\{ A_1, A_2 \right\} }{ A_1 A_2 }
        & = & - \half \frac{z_1 + z_2}{z_1 - z_2}
                \left(
                    \frac{\det Y_1}{\det X_1} +
                    \frac{\det Y_2}{\det X_2} -
                    \frac{\det Y_1}{\det X_1} -
                    \frac{\det Y_2}{\det X_2}
                \right) + \\
        & + & {\tr}_2 (I \otimes  f_{N-1})
              \left(
                \zeta_2 d_2 X^{-1}_2 - \zeta_2 Y^{-1}_2 -
                \kappa_2 d_2 X^{-1}_2 + \kappa_2 Y^{-1}_2
              \right) \\
        & + & {\tr}_1 (f_{N-1} \otimes I)
              \left(
                \zeta_1 d_1 X^{-1}_1 - \zeta_1 Y^{-1}_1 -
                \kappa_1 d_1 X^{-1}_1 + \kappa_1 Y^{-1}_1
              \right)
\end{eqnarray*}
where the non zero terms can be rewritten as
\[
    {\tr}_1 (f_{N-1} \otimes I)
        \left(
            \nu_1 d_1 X^{-1}_1 - \nu_1 Y^{-1}_1
        \right) + \\
    +
    {\tr}_2 (I \otimes f_{N-1})
        \left(
            \nu_2 d_2 X^{-1}_2 - \nu_2 Y^{-1}_2
        \right)
\]
with $\nu \equiv \zeta - \kappa$.

All we have to do now is to decompose $\nu$ over the rows of $Y$
\begin{eqnarray*}
    \left\{
        \begin{array}{lcl}
            \nu_1 & = & \lambda^{(1)}_1 a^{(1)}_1 +
                        \ldots +
                        \lambda^{(1)}_{N-2} a^{(N-2})_1 +
                        \lambda^{(1)}_{N-1} \xi_1 \\
            \nu_2 & = & \lambda^{(2)}_1 a^{(1)}_2 +
                        \ldots +
                        \lambda^{(2)}_{N-2} a^{(N-2})_2 +
                        \lambda^{(2)}_{N-1} \xi_2
        \end{array}
    \right.
\end{eqnarray*}
and notice that only the terms containing $\lambda_{N-2}$ contribute (the
others are either zero or, for the last one, cancel) in order to finally
obtain
\[
    \left\{ \frac{\det X_1}{\det Y_1}, \frac{\det X_2}{\det Y_2} \right\}
        = \frac{\det X_1}{\det Y_1} \frac{\det X_2}{\det Y_2}
          \left(
            \lambda^{(1)}_{N-2} \det Z_1 + \lambda^{(2)}_{N-2} \det Z_2
          \right)
\]
which implies (\ref{eq:AAbrack}).

This terminates our proof that $z_i$ and $w_j$ have the canonical brackets and
means that we have succeeded in constructing the separated variables in a
purely algebraic way.

%% file: reconstr.tex
\section{Reconstruction of $l(z)$}
\label{sec:reconstr}

We have seen that we could relatively easily construct a point on the Jacobian
starting from the matrix elements of $l(z)$. In this section we are going to
show that, conversely, we can reconstruct these matrix elements from a set of
$g$ points lying on the Riemann surface and that this gives us an isomorphism
between the space of matrices and the $\J - \X$.

The reconstruction of $l(z)$ is more difficult because we cannot explicitly
give the matrix elements in terms of the $z_i$ and $w_i$ for $N > 2$. However
we are going to show that we can write a system of equations from which these
elements can be found.

\subsection{Polynomials $X_k$}

Let us start with a point on the Jacobian $(w_i, z_i)_{1 \leq i \leq g}$
satisfying $\psi_1(z_i) = 0$ with $\psi$ as before.

Then we clearly have
\[
    \left\{
        \begin{array}{lcl}
            b(z_i) \phi(z_i) & = & 0 \\
            \left( d(z_i) + w_i \right) \phi(z_i) & = & 0
        \end{array}
    \right.
\]
and this implies that $N$ determinants obtained by dropping $k-th$ row from
the $N \times (N-1)$ matrix made of $b(z_i)$ and $d(z_i) + w_i$ are
null. First, let us consider those of them which contain the first row: for
$k = 1, \ldots, N-1$ we have
\[
    X_k(w_i, z_i) \equiv \det \left(
                        \begin{array}{c}
                            b(z_i)              \\
                            (d(z_i)+w_i)_1      \\
                            \vdots              \\
                            (d(z_i)+w_i)_{k-1}  \\
                            (d(z_i)+w_i)_{k+1}  \\
                            \vdots              \\
                            (d(z_i)+w_i)_{N-1}
                        \end{array}
                   \right) = 0
\]

This determinant can be developed into a sum of monomials $z_i^k w_i^l$ with
$l \leq N-2$. Remembering from (\ref{eq:ldegree}) that all elements
of $b(z)$ are of degree $n-2$ in $z$ except for the last one which is of
degree $n-1$, it is not difficult to see that most of these monomials satisfy
the condition
\begin{equation}
    k \leq (N-1-l)n-2
    \label{eq:condholo}
\end{equation}
The only terms of higher degree in $z$ come from decomposing the determinant
along the first row and taking the last element of it. Such terms will have
the form $z_i^{(N-l-1)n-1}w_i^l$ and appear with coefficient equal to the
leading coefficient of $l_{1N}$ which is (being also the leading coefficient
of $\det l(z)$) a constant, i.e. lies in the center of the Poisson algebra.

Hence, if we move these `exceptional' terms to the RHS of the equations
\[
    X_k(w_i, z_i) = 0 \quad \quad 1 \leq i \leq g
\]
we obtain a system of linear non homogeneous equations for $g$ coefficients of
$X_k$ which we can solve and obtain (almost all) coefficients of $X_k$ in the
form of (\ref{eq:fholo}) because the terms in denominator, by construction,
satisfy the condition (\ref{eq:condholo}). In other words, we have expressed
all $X_k$ in terms of holomorphic functions on the affine Jacobian of $(w_i,
z_i)$.

Next we turn to the remaining (yet unused) determinant
\[
    X_N(w_i, z_i) \equiv \det\left(d(z_i) + w_i\right)_{1 \leq i \leq g}
\]
It contains a term $w^{N-1}$ which doesn't appear in any holomorphic
differential and also, unlike in the previous case, we have here a lot of
terms of the form $z_i^{(N-1-l)n-1} w_i^l$ with rather complicated
coefficients.

However we may remark that, by definition of $d(z)$ as lower bottom part
of $l(z)$, the coefficient of $w^l z^{(N-1-l)n-1}$ in the decomposition of
$X_N$ is the same as the coefficient of $w^{l+1} z^{(N-1-l)n-1}$ in the
decomposition of $\det\left(l(z) + w\right)$. And this coefficient is the
dominant coefficient of the polynomial $t_{N-l-1}$ appearing in the
characteristic polynomial (\ref{eq:def_t}) and so, according to
(\ref{eq:brdet}) and the remark following it, is a constant.

Hence after moving these constant terms to the RHS we can find all the other
coefficients of $X_N$ in terms of the holomorphic functions on the affine
Jacobian in the same way as above.

To summarize, starting from a point $(w_i, z_i)$ on the Jacobian we have
found all $X_k$ for $1 \leq k \leq N$ as holomorphic functions of it.

\subsection{Reconstruction of $l(z)$}

To really reconstruct $l(z)$ we now have to show that we can express its
matrix elements in terms of $X_1(z, w), \ldots, X_N(z, w)$ and also
$\det\left(l(z) + w\right)$.

It is convenient to consider a $N \times (N-1)$ matrix $L'(z)$ defined by
\[
    L'(z) =
            \left(
                \begin{array}{c}
                    z b(z) \\
                    d(z)
                \end{array}
            \right) -
            z^n \left(
                    \begin{array}{ccccc}
                        0 & 0 & 0 & \ldots & l_{1N}^{(0)} \\
                        0 & 0 & 0 &        & 0 \\
                        1 & 0 & 0 &        & 0 \\
                        0 & 1 & 0 &        & 0 \\
                        \vdots &   & \ddots & \ddots & \vdots \\
                        0 & \ldots & \ldots & 1 & 0
                    \end{array}
                \right)
\]
Note that $L(z) = \Ob(z^{n-1})$ by construction as we have removed all the
elements of degree $n$ from it. We shall also denote by $l'_{ij}(z)$ the
polynomial $l_{ij}(z)$ without the term of degree $n$ (if any).

We also introduce another $N \times (N-1)$ matrix
\[
    L(w, z) = 
        w \left(
            \begin{array}{ccccc}
                0 & 0 & 0 & \ldots & 0 \\
                1 & 0 & 0 & \ldots & 0 \\
                0 & 1 & 0 & \ldots & 0 \\
                \vdots &   & \ddots & & \vdots \\
                \vdots &   & & \ddots & \vdots \\
                0 & \ldots & \ldots & \ldots & 1
            \end{array}
          \right)
        +   \left(
                \begin{array}{c}
                    z b(z) \\
                    d(z)
                \end{array}
            \right)
\]
and $N$ $(N-1) \times (N-1)$ matrices $L'_k(z)$ and $L_k(z, w)$ which are
obtained from $L'(z)$ and and $L(w, z)$ respectively by dropping the
$(k+1)$-th row ($1 \leq k \leq N-1$) or the first row for $k = N$.

By definition of $L_k(z, w)$ and $X_k(z, w)$ we have
\begin{equation}
    \label{eq:landx}
    \det L_k(w, z) = \left\{ \begin{array}{ll}
                                z X_k(w, z) & \hbox{if } k < N \\
                                X_N(w, z) & \hbox{if } k = N
                             \end{array}
                     \right.
\end{equation}
As we have already found $X_k$ in terms of the function on the Jacobian, we'd
like to find the matrix elements of $L_k$ now in terms of $X_k$. Of course, it
cannot be done directly from (\ref{eq:landx}) so we are going to start by
extracting the part of the determinant linear in $L'$.

For this we define $V_k$ by
\[
    L_k(w, z) = V_k(z, w) + L_k(z)
\]
or, more explicitly:
\[
    V_k(w, z) =
        \left(
            \begin{array}{cccccc}
                    &   &     &        &        & l_{1N}^{(0)} z^n  \\
                w                                                   \\
                z^n & w                                             \\
                    &   & z^n & w                                   \\
                    &   &     & \ddots & \ddots                     \\
                    &   &     &        & z^n    & w
            \end{array}
        \right)        
        \begin{array}{l}
        \hbox{\scriptsize $k-1$ rows} \\
        \\
        \hbox{\scriptsize $N-k-1$ rows}
        \end{array}
\]
(there is no first row for $k = N$).

Then
\[
    \det L_k(z, w) =
        \det V_k(z, w) \left( 1 + \tr(V_k^{-1}(z, w)L'_k(z)) + \Ob(L'^2) \right)
\]
and we obviously have
\begin{equation}
    \label{eq:detV}
    \det V_k(z, w) = (-1)^N l_{1N}^{(0)} w^{k-1} z^{(N-k)n}
\end{equation}
(there is no $(-1)^N l_{1N}^{(0)}$ factor for $k=N$).

Let us now calculate the trace. We can see from the definition of $V_k$ that
that the inverse matrix has the form
\[
    V_k^{-1}(z, w) =
        \left(
            \begin{array}{ccccc}
                     & w^{-1}                           \\
                     & -w^{-2}z^n & w^{-1}              \\
                w^2z^{-3n}/l_{1N}^{(0)} & & z^{-n} & -wz^{-2n}       \\
                -w z^{-2n}/l_{1N}^{(0)} & & & z^{-n}                 \\
                z^{-n}/l_{1N}^{(0)}
            \end{array}
        \right)
\]
or, more formally:
\[
    \left(V_k^{-1}\right)_{ij} = \left\{
        \begin{array}{ll}
            (-1)^{i+1-j} w^{-(i+2-j)} z^{n(i+1-j)} &
                \quad 1 \leq i \leq k-1, 2 \leq j \leq i+1 \\
            (-1)^{N-1-i} l_{1N}^{(0)} w^{N-1-i} z^{-n(N-i)} &
                \quad k \leq i \leq N-1, j = 1 \\
            (-1)^{j-i-1} w^{j-i-1} z^{-n(j-i)} &
                \quad k \leq i \leq N-1, j \geq i+1 \\
            0 &
                \quad \hbox{otherwise}
        \end{array}
    \right.
\]

Then it is easy to calculate:
\begin{eqnarray*}
    \lefteqn{\tr(V_k^{-1}(z, w)L'_k(z)) = } \\
      & = & \sum_{i=1}^{k-1} \sum_{j=0}^{i-1}
                (-1)^j w^{-j-1} z^{nj} (L'_k)_{i-j+1,i} + \\
      & + & \sum_{i=k}^{N-1} \sum_{j=0}^{N-1-i}
                (-1)^j w^j z^{-n(j+1)} (L'_k)_{i+j+1,i} \\
      & = & \sum_{j=0}^{k-2} (-1)^j w^{-j-1} z^{nj}
                \sum_{i=j+1}^{k-1} (L'_k)_{i-j+1,i} + \\
      & + & \sum_{j=0}^{N-1-k} (-1)^j w^j z^{-n(j+1)}
                \sum_{i=k}^{N-1-j} (L'_k)_{i+j+1,i}
\end{eqnarray*}

Using (\ref{eq:detV}) we can write down the part of $\det L_k(w, z)$ linear in
$L'(z)$ explicitly as
\begin{eqnarray}
    \label{eq:linearpart}
    \lefteqn{\det V_k(z, w) \tr(V_k^{-1}(z, w)L'_k(z)) = } \nonumber \\
      & = & (-1)^{k-2} (L'_k)_{2,k-1} z^{(N-2)n} + w^{k-2} z^{(N-k)n} \pm \nonumber \\
      &\pm& \ldots - \nonumber \\
      & - & ((L'_k)_{22} + (L'_k)_{33} + \ldots + (L'_k)_{k-1,k-1})
                w^{k-3} z^{(N-k+1)n} + \nonumber \\
      & + & ((L'_k)_{21} + (L'_k)_{32} + \ldots + (L'_k)_{k,k-1}) + \nonumber \\
      & + & ((L'_k)_{k+1,k} + \ldots + (L'_k)_{N-1,N-2} + (L'_k)_{1,N-1})
                w^{k-1} z^{(N-k-1)n} - \nonumber \\
      & - & ((L'_k)_{k+2,k} + \ldots + (L'_k)_{N-1,N-3} + (L'_k)_{1,N-2})
                w^{k} z^{(N-k-2)n} \pm \nonumber \\
      &\pm& \ldots + \nonumber \\
      & + & (-1)^{N-k-1} (L'_k)_{1,k} w^{N-2}
\end{eqnarray}

Remember that by definition of $L'_k$ we have
\[
    (L'_k)_{i,j} = \left\{
        \begin{array}{ll}
            z b_{j+1} \pmod{z^n}
                                   & \hbox{if } i = 1 \hbox{ and } k \neq N \\
            l_{i,j+1} \pmod{z^n}   & \hbox{if } i < k \\
            l_{i+1,j+1} \pmod{z^n} & \hbox{if } i \geq k
        \end{array}
    \right.
\]
so all matrix elements of $b(z)$ and $d(z)$ appear in
(\ref{eq:linearpart}) and, remembering (\ref{eq:landx}), we see that we have
almost achieved our goal of expressing the coefficients of $l_{ij}(z)$ in
terms of $X_k$.

To make it more precise, let us introduce yet another notation: define
$L'^{(k)}$ by
\begin{equation}
    \label{eq:detLpk}
    L'(z) = \sum_{k=0}^{n-1} z^k L'^{(k)}
\end{equation}

The simple but important observation we can now make is that we can
determine all coefficients of each $L'^{(k)}$ from (\ref{eq:linearpart}).
Indeed, this equation has a triangular structure and we may first use its
first line for all $k$ to find the elements of the second row, then use the
second line to find the elements of the third row using those of the second
one and so on.

Now let us leave aside the linear part (\ref{eq:linearpart}) of $\det L_k$ for
a moment and consider the entire expression for it. More precisely, let us
study $L^p_k$ defined by
\[
    \det L(w, z) = \sum_{p=0}^{N-1} w^p L^p_k(z)
\]

$L^p_k$ are polynomials of $z$ (only). It is immediate to see from the
definition of $L(w, z)$ that the maximal degree of $z$ in a monomial $w^p z^q$
which can occur in $\det L_k(w, z)$ is $(N - p - 1)n - 1$ as there is always a
minor in which all but one $l_{ij}(z)$ are of maximal degree $n$ containing
$w^p$ (except for $k=N$ and $p=N-1$). So we see that $\deg L^p_k = (N - p -
1)n - 1$.

Moreover, as the dominant coefficient of $L^p_k(z)$ comes from the minor
mentioned above it is an expression linear in elements of $L'^{(n-1)}$ - in
fact, it contains either some $l_{1j}^{(0)}$ multiplied by $l_{i+1,i}$
which is just $1$ or $l_{1N}^{(0)}$ (which is not an element of $L'$)
multiplied by $l_{ij}^{(0)}$ with $i \neq j+1$.

According to the above, we may find $L'^{n-1}$ entirely in terms of $X_k$ (and
so in terms of the functions on the affine Jacobian) - and so the dominant
coefficient of $L^p_k(z)$ can be calculated from them as well. Now let us look
at the next coefficient in it: clearly, for the reasons of homogeneity it may
only contain terms linear in $L'^{n-2}$ or quadratic in $L'^{n-1}$. The latter
have been already calculated and the former can be deduced, again, from
(\ref{eq:linearpart}). Then we continue further with the third coefficient of
$L^p_k(z)$ and so on.

In this way we can find the entire polynomial $L^p_k(z)$ in terms of $X_k$ and
so all the matrix elements of $L(w, z)$ can be expressed as holomorphic
functions on the affine Jacobian because of (\ref{eq:detLpk}).

The only remaining unknown coefficients of $l(z)$ are $l_{k1}(z)$ which don't
appear in $L(w, z)$. We can find them from the
characteristic polynomial. The part of it linear in $l_{k1}(z)$ has the form
\[
    l_{11}(z) w^{N-1} + l_{21}(z) w^{N-2}z^n + \ldots + l_{N1}(z) z^{(N-1)n}
\]
from where we can directly determine the coefficients of $l_{i1}(z)$ in terms
of $r(w, z)$ just as above.

In conclusion, we have shown that it is possible to uniquely determine all
coefficients of the matrix elements of $l(z)$ satisfying (\ref{eq:curve}) and
the conditions on its degree from the $g$ points lying on the Riemann surface
defined by the same equation provided these points don't fall on the
theta divisor. This achieves our proof.

%% file: euler.tex
\section{Calculation of the Euler Characteristic}
\label{sec:euler}

The goal of this section is to compute the $q$-Euler characteristic of the
complex $C_0^*$ defined below. It is the same one as discussed in
\cite{NScohom} and \cite{NSeuler} and we apply the same method (using the
affine model constructed in the first parts of this paper) for its
calculation.

\subsection{Grading}

Let us consider the polynomial ring $\A$ freely generated by all coefficients
appearing in the matrix elements of $l(z)$. We can introduce a grading on this
ring in the following natural way: first, prescribe degree $N$ to $z$ and
degree $N n - 1$ to $w$ (choice of $N$ for the degree of $z$ is different from
the conventions used in \cite{NScohom} where $z$ has the degree $1$, but is
more convenient here to avoid fractional degrees).

Next we remark that the grading of $\det l(z)$ must be equal to $N(Nn-1)$
because it is a polynomial of degree $Nn-1$ in $z$ with dominant coefficient
equal to $1$. So the grading of $l_{21}(z) l_{32}(z) \ldots l_{N\,N-1}(z)
l_{1N}(z)$ is $N(Nn-1)$ and hence we see that we must set
$\deg l_{i+1\,i}(z) = Nn$ and $\deg l_{1N} = N(n-1)$. From further
considerations of homogeneity it is clear that the compatible grading must be
defined by
\[
    \deg l_{ij}(z) = Nn + i - j - 1
\]
Now writing
\[
    l_{ij}(z) = l_{ij}^{(0)}z^{\deg_z l_{ij}} +
                l_{ij}^{(1)}z^{\deg_z l_{ij} - 1} +
                \ldots +
                l_{ij}^{(\deg_z l_{ij})}
\]
we also see that
\[
    \deg l_{ij}^{(\alpha)} = N(n -\deg_z l_{ij} + \alpha) + i - j - 1
\]

We may now easily calculate the character of the free ring $\A$:
\begin{eqnarray*}
    \ch \A & = & \left( \prod_{j=1}^{N-1} \prod_{\alpha=1}^{n-1}
                           {1 \over 1 - q^{N \alpha + N - j}}\right)
                 \prod_{\alpha=1}^{n-1} {1 \over 1 - q ^{N \alpha}} \times \\
      & \times & \prod_{i=1}^{N-1} \left( \left(
                    \mathop{\prod_{j=1}}_{j \neq i}^N \prod_{\alpha=1}^n
                        {1 \over 1 - q^{N\alpha + i - j}}\right)
                    \prod_{\alpha=1}^N
                        {1 \over 1 - q^{N\alpha}}
                 \right)
\end{eqnarray*}
The product on the first line can be written as
\begin{eqnarray*}
    \prod_{\alpha=1}^{n-1} \prod_{j=0}^{N-1} {1 \over 1 - q^{N\alpha+j}} =
        \frac{\left[N-1\right]!}{\left[Nn-1\right]!}
\end{eqnarray*}
with the standard notations
\begin{eqnarray*}
    \left[ N \right]! & \equiv & \left[1\right] \left[2\right] \ldots \left[N\right] \\
    \left[ n \right]  & \equiv & 1 - q^n
\end{eqnarray*}

As for the second part of the product, we have
\begin{eqnarray*}
    \prod_{i=1}^{N-1} \left( \ldots \right) & = &
        \prod_{i=1}^{N-1} \prod_{j=1}^N \prod_{\alpha=1}^n
            {1 \over 1 - q^{N\alpha+i-j}} \\
  & = & \prod_{\alpha=1}^n \prod_{i=1}^{N-1} \prod_{k=-(N-i)}^{i-1}
            {1 \over 1 - q^{N\alpha+k}} \\
  & = & \prod_{i=1}^{N-1} {\left[i-1\right]! \over \left[Nn+i-1\right]!}
\end{eqnarray*}

Combining both parts together we find that
\[
    \ch \A = \prod_{i=1}^N {\left[i-1\right]! \over \left[Nn+i-2\right]!}
\]

\subsection{The quotient ring $\A_0$}

Next object we are interested in is the ring $\F$ freely generated by the
coefficients of the polynomials $t_k(z)$. From their defining equation
(\ref{eq:def_t}) it follows that, for the reasons of homogeneity, the grading
of $t_k(z)$ is $nk - 1$ and its dominant coefficient has grading $N - k$.
Hence
\begin{equation}
    \ch \F = \prod_{k=1}^N \prod_{i=1}^{nk}{1 \over 1 - q^{Ni - k}}
    \label{eq:ch_F}
\end{equation}
where the factor with $k=N$ and $i=1$ is excluded.

As in \cite{NScohom} we define the ring $\A_0$ as
\[
    \A_0 = \A / (\F^{\times} \A), \quad \F^{\times} = \sum_{k,i} \F t_k^{(i)}
\]

And the proposition {\bf 1} of \cite{NSeuler} gives us its character as
\begin{equation}
    \ch \A_0 = {\ch \A \over \ch \F}
    \label{eq:ch_A}
\end{equation}

\subsection{Cohomology groups}

As in the hyperelliptic case we can introduce the differential operators
$D_{ik}$ corresponding to the holomorphic forms on the Riemann surface which
are
\[
    \omega_{ik}(z, w) = {w^{k-1} z^{i-1} \over \partial_w r(z, w)} dz
    \qquad \hbox{with }
    \left\{
    \begin{array}{lcl}
        k & = & 1, \ldots, N-1 \\
        i & = & 1, \ldots, (N - k)n - 1
    \end{array}
    \right.
\]
We make them act on the ring $\A$ by
\[
    D_{ik}x \equiv \left\{t_{N-k+1}^{((N-k+1)n - i - 1)}, x\right\}
\]

With our choice of grading for $z$ and $w$ we have
\[
    \deg D_{ik} = - \deg \omega_{ik} = (Nn - 1)(N - k) - Ni
\]
This grading is, of course, compatible with the action on $\A$:
\[
    D_{ik} \A^{(n)} \subset \A^{(n + \deg D_{ik})}
\]

The character of the ring $\D$ freely generated by all $D_{ik}$ is
easily computed to be
\begin{equation}
    \label{eq:ch_D}
    \ch \D = \prod_{k=1}^{N-1} \prod_{i=1}^{nk-1} {1 \over 1 - q^{Ni-k}}
\end{equation}

We also introduce the differential forms $\tau_{ik}$ dual to the vector fields
$D_{ik}$ and consider the linear spaces $C^k$ for $k = 0, \ldots, g$ spanned
by forms
\begin{equation}
    \label{eq:defCk}
    x_{i_1 \ldots i_k} d\,\tau_{i_1} \wedge \cdots \wedge d\,\tau_{i_k}
\end{equation}
with $x_{i_1 \ldots i_k} \in \A$.

The differential
\[
    d = \sum_{ik} d\,\tau_{ik} D_{ik}
\]
acts from $C^k$ to $C^{k+1}$: as usual, we first apply $D_{ik}$ to the
coefficients $x_{i_1 \ldots i_k}$ and then take exterior product with
$d\,\tau_{ik}$. Thus we have a complex
\[
    0 \rightarrow C^0 \stackrel{d}{\rightarrow} C^1
                      \stackrel{d}{\rightarrow} \cdots
                      \stackrel{d}{\rightarrow} C^g \rightarrow 0
\]

If we fix the degrees of $\tau_{ik}$ to be opposite to $\deg D_{ik}$ the
degree of the differential $d$ is $0$ and so its action descends to the graded
complex
\[
    0 \rightarrow C_0^0 \stackrel{d}{\rightarrow} C_0^1
                        \stackrel{d}{\rightarrow} \cdots
                        \stackrel{d}{\rightarrow} C_0^g \rightarrow 0
\]
where $C_0^k$ are spanned by (\ref{eq:defCk}) with $x_{i_1 \ldots i_k} \in
\A_0$.

\subsection{Euler characteristic}

The formula (\ref{eq:ch_D}) was the last ingredient we needed to compute
\[
    \chi_q(C_0^*) = (-1)^g q^{-\sum_{ik} \deg D_{ik}} \frac{\ch \A_0}{\ch \D}
\]
Using (\ref{eq:ch_A}) we get
\begin{eqnarray}
    \chi_q(C_0^*) & = &
        (-1)^g q^{-\sum_{ik} \deg D_{ik}}
        \prod_{k=1}^N \frac{[k-1]!}{[Nn+k-2]!} \nonumber \\
    & \times & 
        \prod_{k=1}^{N-1} \left( [(Nn-1)k] \prod_{i=1}^{nk-1} [Ni-k]^2) \right)
        \prod_{i=1}^{Nn-1} [Ni]
    \label{eq:q_euler}
\end{eqnarray}

Counting the number of $q$-numbers in the numerator and denominator we see
that it is the same ($N(Nn-1)+N(N-1)/2$) and hence (\ref{eq:q_euler}) admits a
finite limit when $q \to 1$:
\begin{eqnarray}
    \chi(C_0^*) & = &
        (-1)^g (Nn-1)^{N-1} N^{Nn-1}
        \prod_{k=1}^N \frac{(k-1)!}{(Nn+k-2)!} \nonumber \\
    & \times &
        \prod_{k=1}^{N-1} \left( k \prod_{i=1}^{nk-1} (Ni-k)^2 \right)
        (Nn-1)!
    \label{eq:euler}
\end{eqnarray}

Let us now study how does the Euler characteristic depend on $n$:
\begin{eqnarray*}
    {\chi_{n+1} \over \chi_{n}}
        & = & \prod_{k=1}^N \prod_{i=1}^N {1 \over (nN + k - 2 + i)} \\
        & \times & \prod_{k=1}^{N-1} \prod_{i=1}^k ((nk + i - 1)N - k) \\
        & \times & \prod_{k=1}^N \prod_{i=1}^k ((nk + i)N - k)
\end{eqnarray*}
Hence when $n \to \infty$ we have
\[
    {\chi_{n+1} \over \chi_n} \sim e^{2 \sum_{k=1}^{N-1} k \log k + N \log N}
\]
and so we see that $\chi$ as a function of $n$ grows faster than $N^g$ and, a
fortiori, than $2^g$ which should be its asymptotic growth rate if the
highest cohomology functions had a form as simple as in \cite{NScohom}.

We can also note that even for $N=3$ case the value of the Euler
characteristic is already not compatible any more with the conjectures of that
paper and so such simple structure of the highest cohomology group is really
specific to the hyperelliptic case only.

%% file: end.tex
\section{Conclusion}

In this paper we were able to find a new explicit algebraic construction of
the separated variables essentially independent on the concrete integrable
system and which can be extended to the case of quantum integrable models.
Using this affine model we have found a relatively simple expression for the
Euler characteristic of the Jacobian of a spectral curve.

Examining the rate of growth of the Euler characteristic, we realized that
the simple situation of the hyperelliptic case described in \cite{NScohom}
doesn't hold any more in general. This means that we cannot obtain the values
of all observables in our integrable system by using just the functions with
the simple poles of the form (\ref{eq:fholo}) but that we probably need to
consider the the functions with the higher order poles as well.

\section{Acknowledgements}

This work was supported by INTAS grant 00-00055.

%% file: main.bbl
\begin{thebibliography}{}
\bibitem{NScohom}
A. Nakayashiki and F.A. Smirnov,
{\it Cohomologies of Affine Jacobi Varieties and Integrable Systems}.
math-ph/0001017

\bibitem{NSeuler}
A. Nakayashiki and F.A. Smirnov,
{\it Euler Characteristics of Theta Divisors of Jacobians for Spectral Curves}.
math-ag/0012251

\bibitem{Beauville}
A. Beauville,
{\it Jacobiennes spectrales et syst\'emes hamiltoniens compl\'etement
int\'egrables}.
Acta Math. {\bf 164} (1990) 211

\bibitem{Mumford}
D. Mumford,
{\it Tata Lectures on Theta}, vol. I and II, Birkh\"auser, Boston (1983)

\bibitem{Sklyanin}
E.K. Sklyanin,
{\it Separation of Variables in the Classical Integrable $SL(3)$ Magnetic
Chain}.
hep-th/9211126

\bibitem{Scott}
D.R.D. Scott,
{\it Classical Functional Bethe Ansatz for $SL(N)$: Separation of Variables
for the Magnetic Chain}.
hep-th/9403030

\bibitem{Dubrovin}
P. Diener and B.A. Dubrovin,
{\it Algebraic-Geometrical Darboux Coordinates in R-Matrix Formalism}.
SISSA ISAS 88/94/FM

\bibitem{qSmirnov}
F.A. Smirnov,
{\it Separation of variables for quantum integrable models related to
$U_q(\hat{sl}_N)$}.
math-ph/0109013

\end{thebibliography}
